\documentclass[a4paper,12pt]{article}

\usepackage{graphics}
\usepackage{fancyheadings}
\usepackage{pifont}
\usepackage{epsfig}
\input amssym.def

\renewcommand{\theequation}{\arabic{section}.\arabic{equation}}

\begin{document} 

\begin{titlepage}
\begin{center}
\vspace{2cm}
{\LARGE \bf Infrared problems with dimensional regularisation at
finite temperature.}\\

\vspace{2cm}

\centerline{\bf F. Pauquay$^a$}
\vskip 1cm
$^a$Department of Applied Mathematics and Theoretical Physics,\\
Centre for Mathematical Sciences, University of Cambridge, Cambridge CB3 0WA \\
         f.h.r.pauquay@damtp.cam.ac.uk  
{\large }

\vspace{3cm}

{\bf Abstract}\\
\bigskip
\end{center}
We address the question whether we have to perform the
analytic continuation to $4$ space-time dimensions before or after the
perturbative expansion has been made. Using the simple model of large-$N$
scalar field theory, we show how this affects quantities like the
thermal mass and the pressure, why the two procedures
give different results, and how this relates to the infrared behaviour of
the theory. We conclude that the correct procedure is to take the
limit $d\rightarrow 4$ before performing the perturbative expansion. 
Infrared divergences arise if one makes the perturbative expansion
before going to the physical space-time dimension and they correspond to
logarithms of the coupling in the exact (nonperturbative or resummed) expression. 
Finally, we outline how this may be relevant to the infrared
problems encountered in thermal QCD.

\end{titlepage}

\section{Introduction}
It has been known for a long time that beyond-leading-order
calculations in massless thermal field theories exhibit infrared divergences \cite{lebellac,Landsman}.
In the case of abelian theories, systematic
resummation techniques have been developed to cure infrared
divergences. This leads to perturbative expansions which are nonanalytic in the coupling.
In contrast to the case of abelian theories, no resummation  
method has so far managed to cure infrared divergences in non-abelian
gauge theories like QCD \cite{lebellac}.
In QED, infrared divergences mainly arise in Feynman
diagrams with multiple insertions of the static photon
self-energy.
A resummation of all diagrams with an arbitrary number of
such insertions (ring diagrams) can be shown to eliminate all infrared
divergences. Consequently, quantities like the free energy can, in principle, be
expanded to arbitrary order in powers of the coupling. In QCD however, since gluons couple to
themselves, the behaviour of the theory is
much more singular: diagrams  with static spatial gluon loops are
increasingly infrared divergent when their order gets higher.  
This may signal the presence of a magnetic mass which would act as an
infrared cut-off to regulate infrared divergent
integrals. Although such a magnetic mass is  potentially
of order $g^2\,T$ since spacelike gluons remain massless to lowest
order, it cannot be computed perturbatively. This is because beyond one-loop order,
all the diagrams in a self-consistent expansion of the magnetic mass will be of the same order in the
coupling \cite{Linde}. As a result, unresummed perturbation theory does not seem
to work any more and all those diagrams that contribute to the same
order in the coupling should somehow be resummed in a similar fashion to simpler theories
like $\lambda\phi^4$ or QED. Due to the complexity of
QCD, this is unfortunately impossible in practice.\\
If one uses dimensional regularisation, with $d$ the dimension of
space-time, the breakdown of perturbation theory from
some order signals itself through infrared $1/(d-4)$ poles in the
perturbative series from a given order. It reflects the fact that the
exact quantity one tries to compute is nonanalytic in the coupling
when $d=4$. In this work we therefore examine dimensional
regularisation at finite temperature in more detail.  
Dimensional regularisation was devised as a method to separate the
finite and divergent parts of Feynman diagrams arising beyond leading
order in perturbation theory \cite{pesk}. 
At finite temperature, there are infrared $1/(d-4)$ divergences in addition to
the zero temperature ultraviolet ones so, if one use dimensional
regularisation in the usual way, one has to make sure that these IR
divergences somehow cancel each other in the expression for physical quantities. 
Consequently, one could wonder whether making an expansion in powers
of $g$ and $d\rightarrow 4$ commute. This is relevant because if we keep $d-4$
nonzero, the perturbative expansion in powers of $g$ actually involves
$g/(d-4)$ and therefore is illegal since we are due to take $d$ to
$4$ at the end of the calculation. As a result, an expansion in powers
of $g$ is is only valid if $1/(d-4)$ poles somehow cancel each other in the expression for
physical quantities. A case where the previous limits do not commute
is the following: consider 
\begin{equation} A = \frac{C}{d-4+g^2} \label{example}\end{equation}
If we first expand $A$ in power series of the coupling, its terms are
divergent in the limit $d\rightarrow 4$ :
\begin{equation} A =\frac{C}{d-4}-\frac{C\,g^2}{(d-4)^2}+... \end{equation}      
However, in $4$ dimensions, $A=C/g^2$ is perfectly finite and is
nonanalytic in the coupling.\\   
This work investigates the consequences of this kind of problem in
the case of large-$N$ scalar theory, which will give us some insight
about the way things might work in QCD.  
In next two sections, we derive and renormalise an exact formula for
the thermal pressure in the case of large-$N$ scalar field theory. In
section 4, we study the singular structure of the expansion of the
thermal mass and show that infrared divergences arise from a given
order onwards in the renormalised perturbation series.
We argue that those divergences crop up because the perturbative
expansion has been performed before taking $d$ to the physical
dimension, in the same fashion as for the example given in
(\ref{example}). We will see that the way to rescue the situation is
to resumm the perturbative series and that we then obtain the correct
expansion, ie the one we get if we go to $4$ space-time dimensions first.
Finally, we show this problem originates in the fact that
the finite temperature contribution to the self-energy is actually nonanalytic in the
coupling when $d=4$ and involves terms proportional to $g\log g$ and
half-integer powers of the coupling.  

\section{Unrenormalised formula}
\setcounter{equation}{0}

In this work, we use large-$N$ scalar field theory as a toy model to
investigate the commutativity of the analytic continuation to four space-time
dimensions and the perturbative expansion. Consider the
theory described by the unrenormalised Lagrangian density 
\begin{equation}
{\cal L}(\lambda,x) = {1\over 2} (\partial_\mu \phi)^2 - {1\over 2}m^2\,\phi^2- \frac{\lambda}{4!}\frac{3}{N+2} \,\phi^4 
\end{equation}
where $\phi$ is an $N$-component field. We are interested in the limit
$N\rightarrow \infty$ with $\lambda/N$ fixed.
The grand canonical partition function reads
\begin{equation}
Z_{\beta}(\lambda) = \int{\cal D}\phi \;\exp\left[i\int_{\tau}^{\tau-i\beta} dx^0 \int
d^{d-1}x\; {\cal L}(\lambda,x) \right]   
\end{equation}
where $\beta=1/T$ and $\tau$ is some arbitrary time.
The thermodynamic definition of the pressure is, as usual, 
\begin{equation} P=T\,V^{-1}\log Z_{\beta} \end{equation} 
If we perform the change of variables $\phi\rightarrow
\lambda^{1/2}\phi\;$ in the path integral, we can make the coupling
dependence of the partition function more apparent \cite{trace} and obtain 
\begin{equation}
\frac{Z_{\beta}(\lambda)}{Z_{\beta}(0)} = \frac{\int{\cal D}\phi
\;\exp\left[\frac{i}{\lambda}\int_{\tau}^{\tau-i\beta} dx^0 \int
d^{d-1}x\; {\cal L}(1,x) \right]}{\int{\cal D}\phi
\;\exp\left[\frac{i}{\lambda}\int_{\tau}^{\tau-i\beta} dx^0 \int d^{d-1}x\; {\cal L}(0,x) \right]} \label{ZdivZ}\end{equation}
If we define $\hat{P}=P(T,\lambda)-P(T,0)$ and use the
translation invariance property that 
\begin{equation} 
\left <\int_{\tau}^{\tau-i\beta}dx^0 \int d^{d-1}x\;{\cal L}(1,x)\right>=-i\beta\,V \left<{\cal L}(1,0)\right > \end{equation}
it is straightforward to obtain from (\ref{ZdivZ}) the unrenormalised equation 
\begin{equation}
\frac{\partial\hat{P}(T,\lambda)}{\partial \lambda}=-\frac{1}{\lambda}\left(\left<{\cal L}(\lambda,0)\right>-\left<{\cal L}(0,0)\right>_{FREE}\right) \label{unr} \end{equation}
Because it will simplify the equations in what follows we will
hereafter use the notation $\Delta A \equiv <A>-<A>_{FREE}$ for any operator $A$.

\subsection{Calculation of $\,\Delta{\cal L}$ }
\setcounter{equation}{0}

The main analytic non-perturbative tools available here to compute 
$\,\Delta{\cal L}$ are the
Schwinger-Dyson equations which can be regarded as the quantum
equations of motion for Green's functions \cite{pesk}. These equations are the
same as at zero temperature; the only difference lies in the boundary
conditions : at finite temperature, the time periodicity (for bosons) or anti-periodicity (for
fermions) of classical fields (ie the integration variable in the path
integral) provides Green's functions with temperature
dependent boundary conditions. 
In large-$N$ scalar field theory, the self-energy is exactly
calculable and it is therefore desirable to derive an expression for $\Delta{\cal L}$ that only involves the propagator
since, using the Dyson resummation, the latter can in turn be written in terms of
the self-energy.
To achieve this, we have to express the thermal four-point function in $\Delta{\cal L}$ in terms of the propagator. 
This can be done using the Schwinger-Dyson equation for the propagator.
To derive this equation, consider the generating functional of the thermal Green's functions :
\begin{equation} Z_{\beta}(J)=\int\,{\cal D}\phi \exp\left[ i \int_C d^d x \left( {\cal L}(\phi)+\phi(x)\,J(x) \right) \right]  \label{genfunc} \end{equation}
We use the Keldysh variant of the real-time thermal field theory \cite{lebellac,L1}, so the contour $C$ runs along the real axis from $-\infty $ to $\infty $, back to $-\infty $ and then down to $-\infty - i\beta$.
According to the usual Feynman prescription, we have to add
$i\epsilon\phi^2$ to the Lagrangian density in the generating
functional (\ref{genfunc}) in order to ensure
the convergence of the path integral for large $\phi$ (in what follows, this
prescription will be implicit). 
Then we have the following identity
\begin{equation} \int{\cal  D}\phi \frac{\delta}{\delta\,\phi(x)}\;\exp\left[ i \int_C d^d x' \left( {\cal L}(\phi)+\phi(x')\,J(x') \right) \right]=0   
\end{equation}
which reads 
\begin{eqnarray}\nonumber \int {\cal D}\phi \; \left[
(\partial^2 +m^2 )\phi(x))+
\frac{\lambda}{2}\frac{1}{N+2}(\phi(x))^3\right] \;\;\;\;\;\;\;\;\;\;\;\;\;\;&&\\ \exp\left[ i \int_C d^d x' \left( {\cal L}(\phi)+\phi(x')\,J(x') \right) \right]&=&  Z_{\beta}(J)\; J(x)
\end{eqnarray}
By applying $ \delta/\delta \,\phi(y) $ to both sides of this equation and setting $J$ to zero, we obtain 

\begin{equation}
\left < T_c\,\phi(y)(\partial^2 +m^2 )\phi(x)\right > + \frac{\lambda}{2}\frac{1}{N+2} \left <T_c\, (\phi(x))^3 \phi(y)\right > = -i\, \delta_c^{(d)}(x-y)
\label{DSequ}
\end{equation}
and similarly
\begin{equation} \left < T_c\,\phi(y)(\partial^2 +m^2
)\phi(x)\right >_{FREE}= -i\, \delta_c^{(d)}(x-y) \end{equation}
where $T_C$ and $\delta_c$ denote the time ordering and delta function along the integration contour in complex time.
When $x=y$, the last two equations lead to 
\begin{equation}
\left<\phi\,(\partial^2 +m^2 )\,\phi\right>+\frac{\lambda}{2}\frac{1}{N+2}\left<\phi^4 \right>= \left< \phi\,(\partial^2 +m^2 )\,\phi\right>_{FREE} \label{DS2} \end{equation}
This implies

\begin{equation}
\Delta{\cal L}= -\frac{1}{4}\,\Delta\left[ \phi\,(\partial^{2}+m^2)\,\phi \right]\end{equation}
With the Keldysh contour, we get 
\begin{equation}
\Delta{\cal L}= \frac{1}{4}\int\frac{d^dq}{(2\pi)^d}\;(q^2-m^2)\left(D_{12}(q)-D_{12,FREE}(q)\right) \end{equation}
where the subscript 12 refers to the element of the $2\times2$ thermal matrix propagator.
This propagator has the structure

\begin{equation}
{\bf D}(q)={\bf M}(q^0) \ \tilde{\bf D}(q) \  {\bf M}(q^0) \label{mstruct}\end{equation}
with
\begin{eqnarray}\nonumber 
{\bf M}(q^0) &=& \sqrt{n(q^0)} \left[\matrix{ e^{{1\over 2} \beta |q^0|} & 
e^{-{1\over 2}\beta q^0} \cr e^{{1\over 2}\beta q^0} & e^{{1\over 
2}\beta |q^0|}\cr}\right ]\\ \nonumber \\
\nonumber \tilde{\bf D}(q) &=&   \left[ \matrix{D(q)&0\ \cr
                   0&  D^*(q)\cr}\right ]\\ \nonumber && \\
 D(q)&=&{i\over {q^2-m^2-\Pi(q,m,T)+i\epsilon}} 
\end{eqnarray}
where $n(q^0)$ is the Bose distribution
$\left(e^{\beta\,|q^0|}-1\right)^{-1}$.
This implies 
\begin{equation}
D_{12}(q)=-2\,\epsilon(q^0)\,n(q^0)\,{\rm Im} \left(\frac{1}{q^2-m^2-\Pi(q,T,m)+i\epsilon}\right)\end{equation}
In the case of large-$N$ scalar theory, the self-energy $\Pi(q,m,T)$ is real and independent of the momentum.
In consequence, we have
\begin{equation}
{\rm Im} \left(\frac{1}{q^2-m^2-\Pi+i\epsilon}\right)= -\pi\,\delta(q^2-m^2-\Pi(m,T)) \end{equation}
Hence  we find
\begin{equation} 
\Delta{\cal L}=\frac{\Pi(m,T)}{2}\,M_T(m^2+\Pi(m,T))
\label{dL}\end{equation}
where $M$, $N_T$ and $M_T$ are defined as :
\begin{equation}
M(m^2)=\frac{1}{2}\int\frac{d^dq}{(2\pi)^d} \frac{i}{q^2-m^2+i\epsilon}=\frac{1}{2}\int\frac{d^dq}{(2\pi)^d} 2\pi\,\delta^+(q^2-m^2)\label{M1}\end{equation}
\begin{equation} N_T(m^2)=\int {d^dq\over(2\pi)^d}\,2\pi\,\delta ^+(q^2-m^2)\, n(q^0)\end{equation}
\begin{equation} M_T(m^2)=M(m^2)+N_T(m^2)  \label{MT}\end{equation}
To make use of (\ref{dL}), we need to express all the bare parameters of the Lagrangian in terms of their renormalised counterparts.

\section{Renormalisation}
\setcounter{equation}{0}

We choose a renormalisation scheme that renders $\Delta{\cal L}$ as
simple as possible. Since we are dealing with exact equations, it is
suitable to define the renormalised mass through the Dyson equation,
which is valid to all orders. 
At $T=0$ and in the on-shell renormalisation scheme, define the renormalised
mass as
\begin{equation}m_R^2=m^2+\Pi(q^2=m_R^2,m,T=0) \end{equation}
For large-$N$ scalar theory, the Dyson equation reads
\begin{equation} \Pi(m,T=0)=\lambda M(m_R^2) \end{equation}
so we have
\begin{equation}
m_R^2=m^2+\lambda M(m_R^2) \label{zeroren}\end{equation}
since $N_T$ vanishes at $T=0$.
The integration in (\ref{M1}) gives 
\begin{equation} M(m_R^2)=\frac{\Gamma(1-d/2)}{2(4\pi)^{d/2}}\;m_R^{d-2}\label{M}\end{equation}
At nonzero temperature, it is convenient to define a thermal mass
\begin{equation} \delta m_R^2=\lambda(M_T(m_R^2+\delta m_R^2)-M(m_R^2))\label{rentemp} 
\end{equation}
so (\ref{zeroren}) implies
\begin{equation}
  m_R^2+\delta m_R^2=m^2+\lambda  M_T(m_R^2+\delta m_R^2) 
\label{mr+dmr}\end{equation} 
Comparing (\ref{mr+dmr}) with (\ref{zeroren}), we see that the
definition (\ref{rentemp}) is a natural one.
In what follows, we shall be interested in the case of massless
theories ($m=0$ and therefore $m_R=0$) because there one encounters the infrared problems of thermal perturbation theory.
Since  $M(0)=0$, (\ref{rentemp}) reduces to the following integral equation for the thermal mass 
\begin{equation}
 \delta m_R^2=\lambda  M_T(\delta m_R^2) \label{unrthm}\end{equation} and we obtain
\begin{equation} \Delta{\cal{L}}=\frac{1}{2}\,\lambda\;M_T^2(\delta
m_R^2)=\frac{1}{2}\,\frac{\delta m_R^4}{\lambda}  \label{deltaL}\end{equation}
Beside mass renormalisation, we also have to renormalise the coupling
constant: because $\lambda$ has dimension $M^{4-d}$, we introduce a
renormalisation scale $\mu$ and define a dimensionless renormalised
coupling by 
\begin{equation} \lambda=\mu^{4-d}\,\lambda_R\, Z(\lambda_R)  \label{ren}\end{equation}
where $Z(\lambda_R)$ is a combination of wave-function and vertex renormalisation factors.
It is a function of $\lambda_R$ only because it is dimensionless.
In the case of massive theories, one usually uses the on-shell
renormalisation scheme
\begin{equation}
\lambda_R=\lambda_R(\mu)=\mu^{d-4}\;\lambda+\lambda\;\lambda_R\;M'(m_R^2) \end{equation}
where the prime denotes differentiation of $M$ with respect to its argument.
However, if we consider massless theories, this definition is not useful since $M'(0)$ is infrared divergent.
Instead, define
\begin{equation}
\lambda_R(\mu)=\mu^{d-4}\;\lambda+\lambda\;\lambda_R(\mu)\;M'(\mu^2) \label{scheme}\end{equation}
With (\ref{M}) and (\ref{ren}), this gives 
\begin{equation}
Z(\lambda_R)=\left(1-\frac{\lambda_R\;\Gamma(3-d/2)}{(4-d)\;(4\pi)^{d/2}}\right)^{-1}
\label{Z}\end{equation} 
In fact, the renormalisation scheme defined by (\ref{scheme}) is the
modified minimal subtraction scheme ($\overline{\hbox{MS}}$) since (\ref{Z}) has the expansion 
\begin{equation}
Z=\frac{1}{1+\frac{\lambda_R}{16\pi^2}\left(\frac{1}{d-4}+2\,(\gamma-\log
4\pi)\right)}+{\cal O}(d-4) \end{equation}
If we introduce the $\beta$-function as usual as   
\begin{equation} \beta(\lambda_R)=\mu
\left.\frac{\partial\lambda_R}{\partial\mu}\right|_\lambda
\end{equation} 
and if we differentiate the left and right-hand sides of (\ref{ren})
with respect to $\mu$ at fixed unrenormalised coupling, we get
\begin{equation} 0=4-d+\beta(\lambda_R)\left(\frac{1}{\lambda_R}+\frac{d}{d \lambda_R}\log Z(\lambda_R)\right )\label{difmu}\end{equation}
From this equation and from (\ref{Z}),
\begin{equation} \beta(\lambda_R)= \lambda_R\;(d-4) + \frac{\lambda_R^2\;\Gamma(3-d/2)}{(4\pi)^{d/2}}\label{betaf} \end{equation}
We can also differentiate (\ref{ren}) with respect to $\lambda_R$ at
fixed $\mu$ and then use (\ref{difmu}) to obtain 
\begin{equation} \frac{d\lambda}{\lambda}=d\lambda_R\left[\frac{d}{d \lambda_R}\log Z(\lambda_R)+\frac{1}{\lambda_R}\right]=\frac{d-4}{\beta(\lambda_R)}\,d\lambda_R \label{diflambda}\end{equation}
The pressure being a physical quantity, it is the same before and after renormalisation: $P_R(T\,,\lambda_R)=P(T\,,\lambda)$.
We can therefore combine (\ref{deltaL}), (\ref{Z}) and (\ref{diflambda}) to turn
(\ref{unr}) into the renormalised formula
\begin{equation}
\frac{\partial}{\partial\lambda_R}\hat{P}_R(T,\lambda_R)=\frac{4-d}{\beta(\lambda_R)}
\;\Delta{\cal L}=-\frac{\mu^{d-4}}{2\lambda_R}\,\delta m_R^4 \label{renf} \end{equation}
This equation is of course consistent with its unrenormalised
counterpart (\ref{unr}) since $\lambda=\mu^{4-d}\,\lambda_R$ to
lowest order and $\hat{P}_R=\hat{P}$. 
It also implies that the question whether we get the same answer for
the thermal pressure if use perturbation theory before or after we take the
$d\rightarrow 4$ limit reduces to how $\delta m_R^4$
behaves depending on the chosen procedure. In the next section, we will find
that if we make an expansion in power series of the coupling (ie use standard
diagrammatic Feynman perturbation theory) in $d$ space-time dimensions, not only we get uncancelled
$1/(d-4)$ divergences from order $\lambda_R^4$ onwards but even the finite
second order term is incorrect when we compare it to the exact
alternative expansion we obtain in $4$ dimensions.

\section{The thermal mass}
\setcounter{equation}{0}
We now study the structure of the mass shift when the coupling
constant is chosen as expansion parameter either before or after we go to $4$ dimensions.
Our aim is to  see what kind of expression we get in a perturbative
calculation of the thermal mass and how resummation works to remove infrared
$1/(d-4)$ terms and obtain the unrenormalised non perturbative equation (\ref{unrthm}).
\subsection{Naive perturbative expansion}
In scalar field theories, it is known that the dominating infrared
contributions to the self-energy come from the so-called ring diagrams
and their extension called super-daisy or Hartree-Fock diagrams (see
figure 1).
\begin{figure}[h] 
\epsfig{file=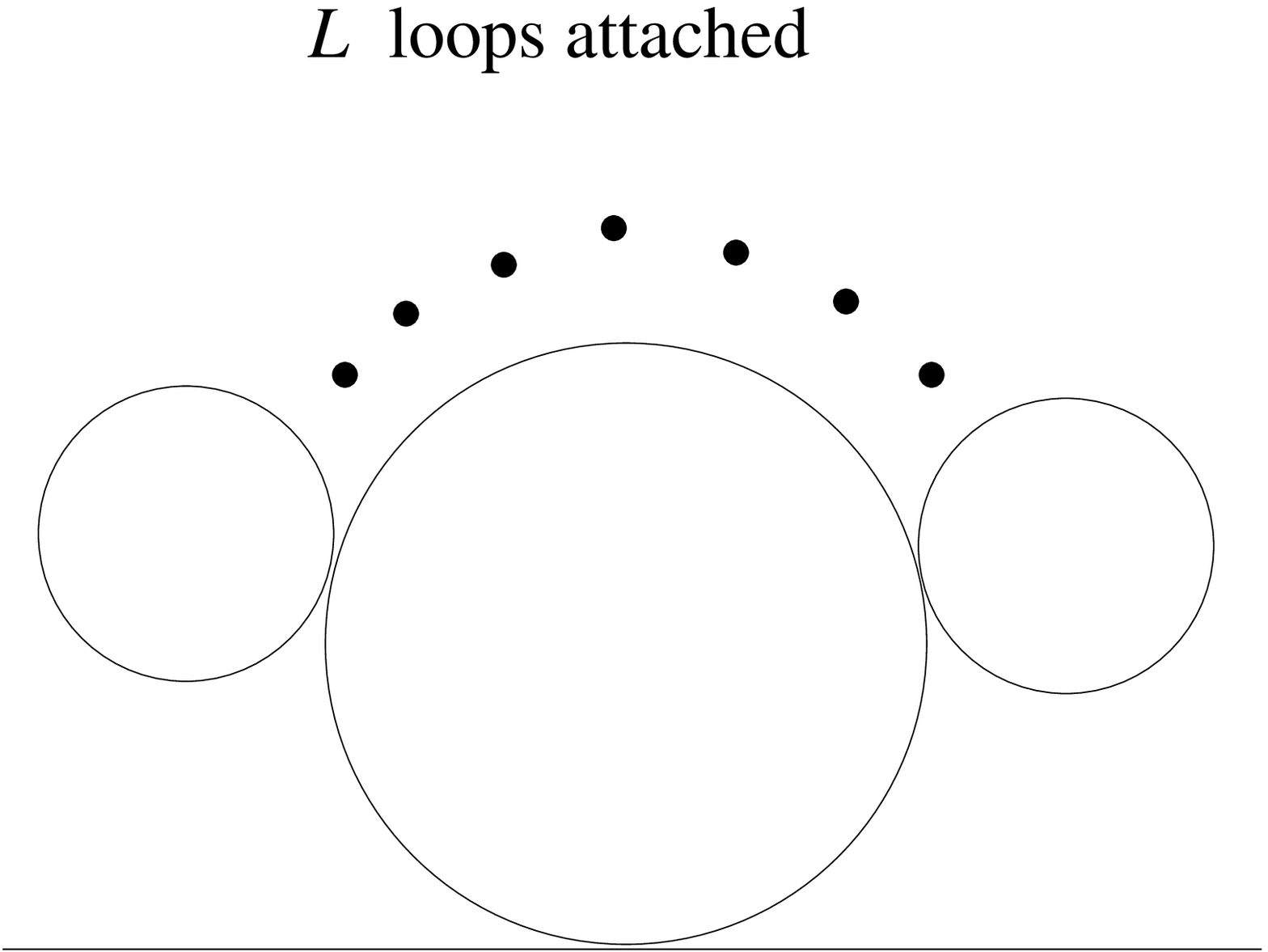,width=6.5cm}
\hspace{1.5cm}
\epsfig{file=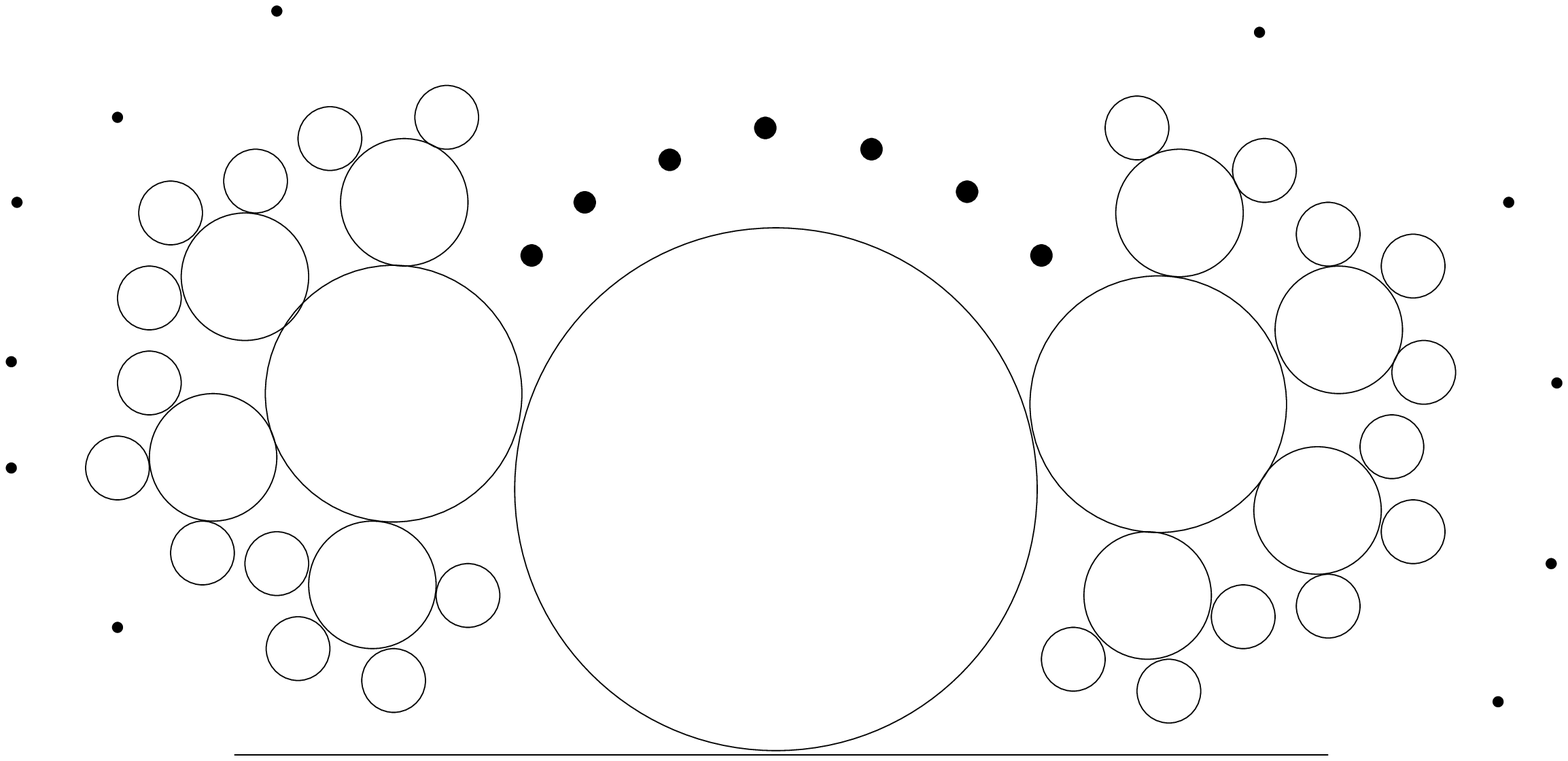,width=7cm}
\caption{Ring and Hartree-Fock Diagrams}
\end{figure}\\
First compute the value of a ring diagram with $L$ external
loops attached using the imaginary-time formalism :
\begin{equation}
\Pi^{(L)}(d,T)=\frac{\lambda T}{2}\; 
m_T^{2L}(-1)^L\left(2\sum_{n=1}^{\infty}\int\frac{d^{d-1}p}{(2\pi)^{d-1}}\frac{1}{(p^2+\omega_n^2)^{L+1}}+
\int\frac{d^{d-1}p}{(2\pi)^{d-1}}\frac{1}{p^{2L+2}}\right)
 \label{piN} \end{equation}
where $\omega_n$ denote the Matsubara frequencies and $m_T \equiv
M_T(0)^{1/2}$ the one-loop thermal mass.
According to the rules of dimensional regularisation, the last term
in this expression, the static mode contribution, vanishes since the integrand does not contain any
dimensionful quantity. 
However, the sum over $L$ of the static mode contributions is 
\begin{equation}  \sum_{L=0}^{\infty} \frac{\lambda T}{2} m_T^{2L}
(-1)^L \int\frac{d^{d-1}p}{(2\pi)^{d-1}}\frac{1}{p^{2L+2}}= \frac{\lambda T}{2}
\int\frac{d^{d-1}p}{(2\pi)^{d-1}}\frac{1}{p^2+m_T^2}=\frac{\lambda T\; m_T}{8\pi}     \label{static}\end{equation}
The perturbative expansion of the right hand side of
(\ref{static}) is inappropriate since it is nonanalytic in the coupling (it
is proportional to $\lambda^{3/2}$) and naive use of dimensional regularisation fails here. 
We can therefore conclude that static mode contributions only make
sense when they have been resummed.\\ 
We now turn to the non-static modes.
In the spirit of dimensional regularisation,
integrals are initially computed in a dimension where they converge
and then analytically continued to four space-time dimensions.
In $d$ dimensions, the non-static part of the self energy is 
\begin{equation} 
\Pi_{n.s.}^{(L)}(d,T)=\lambda T\; 
m_T^{2L}(-1)^L
\frac{(2 \pi T)^{d-2L-3}}{(4\pi)^{\frac{d-1}{2}}}\;\frac{\Gamma(L+1-\frac{d-1}{2})}{L!}\;\zeta(2L+3-d)\label{nspiN} \end{equation}
where $\zeta$ is the standard Riemann zeta-function.
As usual, we get to lowest order
\begin{equation} m_T^2=-\Pi_{n.s.}^{(0)}(4,T)=\frac{\lambda T^2}{24} \end{equation}
Since $\zeta(z)$ is analytic in the whole complex plane except for
$z=1$ where it has a simple pole, $\Pi^{(L)}_{n.s.}(d,T)$ is finite for all
values of $L$ when $d\rightarrow 4$, except for $L=1$. 
In the latter case, we have 
\begin{equation} \Pi_{n.s.}^{(1)}(d,T)     =
\frac{\lambda\,m_T^2}{16\pi^2\,(d-4)}-\frac{\lambda\,m_T^2}{16\pi^2}
\left(\frac{\gamma}{2}-\frac{1}{2}\log 4\pi-\log T\right) +{\cal O}(d-4)\label{pi1}   \end{equation}
It means we still get a $\lambda^2/(d-4)$ pole even if we
have resummed all the static contributions.\\
Since it is known that resumming all ring diagrams provides a finite
result \cite{altherr}, the resummation of all the $\Pi^{(L)}(d,T)$ for $L>1$
should cancel out the $1/(d-4)$ pole coming from  $\Pi^{(1)}(d,T)$ :
\begin{equation}
\sum_{L=2}^{\infty}\Pi^{(L)}_{n.s.}(d,T)=\frac{\lambda
T}{(4\pi)^{\frac{d-1}{2}}}\sum_{n=1}^{\infty} (2\pi n
T)^{d-3}\sum_{L=2}^{\infty}\left(\frac{m_T}{2\pi n T}\right)^{2L}\frac{(-1)^L \Gamma(L-\frac{d-3}{2})}{L!}   \end{equation}
If we use the Taylor expansion 
\begin{equation} 
(1+x)^{\frac{d-3}{2}}=1+\left(\frac{d-3}{2}\right)\;x+\sum_{L=2}^{\infty}\;\frac{1}{L!}\;\frac{\Gamma\left(\frac{d-1}{2}\right)}{\Gamma\left(\frac{d-1}{2}-L\right)}\;x^L
\end{equation} 
and the identity
\begin{equation} \Gamma(x)\;\Gamma(1-x)=\frac{\pi}{\sin(\pi\,x)} \end{equation}
we find, by summing over $n$,
\begin{eqnarray} \nonumber
\sum_{L=2}^{\infty}\Pi_{n.s.}^{(L)}(d,T)=\frac{\lambda\,T\,\pi}{(4\pi)^{\frac{d-1}{2}}\,\Gamma\left(\frac{d-1}{2}\right)}\left[-\sum_{n=1}^{\infty}(\omega_n^2+m_T^2)^{\frac{d-3}{2}}+(2\pi\,T)^{d-3}\,\zeta(3-d)
\right. \\
\left.+\left(\frac{d-3}{2}\right)\,m_T^2\,(2\pi\,T)^{d-5}\;\zeta(5-d)\right]+{\cal
O}(d-4)        \end{eqnarray} 
By expanding this expression around $d=4$, we obtain 
\begin{eqnarray} \nonumber
\sum_{L=2}^{\infty}\Pi_{n.s.}^{(L)}(d,T)= \lambda\,M_T(m_T^2)+\frac{\lambda\,T}{8\pi}\,m_T-m_T^2-\frac{\lambda\,m_T^2}{16\pi^2\,(d-4)} \,\,\,\,\,\,\,\\ \,\,\,+\frac{\lambda\,m_T^2}{16\pi^2}
\left(\frac{\gamma}{2}-\frac{1}{2}\log 4\pi-\log T\right) +{\cal
O}(d-4)             \end{eqnarray} 
As expected, resummation indeed removes the pole: it is cancelled out
by the rest of the perturbative series.
Hence, the resummed result is merely the one-loop self-energy diagram in which the free
propagator has been replaced by the first-order dressed one:
\begin{equation} 
\sum_{L=0}^{\infty}\Pi^{(L)}(d,T)=\lambda\,M_T(m_T^2)+{\cal O}(d-4)\label{sumring}   \end{equation}
Because it comes from the Matsubara sum, it is not clear whether the
$\lambda^2/(d-4)$ divergence was an infrared or an ultraviolet one.
In order to find out, it is interesting to study $\Pi^{(1)}(d,T)$ in the
real-time formalism: this is done in appendix A.1.\\  
In this section, we have so far only been dealing with quantities
involving the unrenormalised coupling so we still need to renormalise
(\ref{unrthm}) to make sure we have properly taken care of UV
divergences before jumping to conclusions about possible remaining divergences.
Expressing bare quantities in terms of renormalised ones,
(\ref{unrthm}) becomes the following integral equation :
\begin{equation} \delta m_R^2=\mu^{4-d}\;\lambda_R\;\left[\tilde{M}(\mu^2,\delta m_R^2)+N_T(\delta m_R^2)\right]\label{rendm}\end{equation}
where $\tilde{M}(\mu^2,\delta m_R^2)\equiv M(\delta m_R^2)-\delta
m_R^2\;M'(\mu^2)\, $ is finite when $d=4$. 
This equation is exact and therefore corresponds to the full summation of
all the Hartree-Fock diagrams.
By studying the ring diagram contributions to the thermal mass
in the ITF and RTF, we have seen that an expansion of $N_T(\delta m_R^2)$ in power series of
$\delta m_R^2 $ gives rise to infrared divergences of the form
$\delta m_R^2/(d-4)$ although $N_T$ itself is perfectly finite when
$d=4$.    
This is because $N_T$ is nonanalytic in $\delta m_R^2 $  at $\delta m_R^2=0 $
(namely, it involves terms which are proportional to $\delta m_R\log\delta m_R$) and so the
Feynman perturbative expansion in powers of $\lambda_R$ is illegal since $\delta m_R \sim
\lambda_R$ to lowest order. In the next section, we investigate
the exact form of $N_T$ and its nonanalyticity in the coupling. 
This will lead us to
the conclusion that the  $\delta m_R^2/(d-4)$  term in the
perturbative expansion of $N_T$ is effectively replaced with $\delta
m_R^2\,\log(\delta m_R^2)$ in the corresponding exact expression (see (\ref{exNT})). 
In section 4.3, we shall see that the divergent term in a perturbative
expansion of $N_T(\delta m_R^2)$ gives rise to infrared divergences in
the renormalised perturbative series from order $\lambda_R^4$ onwards. 

\subsection{Alternative expansion of $N_T$}  
In order to study its dependence in the the coupling, we will now expand $N_T$ in another and more
sensible way than the naive perturbative expansion \cite{Howard}: 
\begin{eqnarray}\nonumber
N_T(m^2)&=&\frac{1}{2}\int\frac{d^{d-1}p}{(2\pi)^{d-1}}\,\frac{p^{d-2}}{\sqrt{p^2+m^2}}\frac{1}{\exp({\beta\sqrt{p^2+m^2}})-1}\\&=&
\frac{T^2}{(4\pi)^{\frac{d-1}{2}}\Gamma(\frac{d-1}{2})}\,\int_{0}^{\infty} du\,\frac{u^{d-2}}{\sqrt{u^2+y^2}}\,\sum_{k=1}^{\infty}\exp{(-k(u^2+y^2)^{1/2})}   \label{NT}                 \end{eqnarray} 
where $y\equiv \beta m$.
We evaluate the sum using the Mellin summation formula (see 
appendix A.2)
\begin{equation} \sum_{k=1}^{\infty}
f(k)=\frac{1}{2i\pi}\,\int_{C-i\infty}^{C+i\infty} dz\,
\zeta(z)\,\phi(z) \end{equation} where $\phi(z)$ is the Mellin
transform of $f(k)$ and $C$ is such that all the poles of the
integrand are to the left of the vertical integration line.
We obtain
\begin{equation}
N_T(m^2)= \frac{T^2}{2i\pi(4\pi)^{\frac{d-1}{2}}\Gamma(\frac{d-1}{2})}\,\int_{C-i\infty}^{C+i\infty} dz\,
        \int_{0}^{\infty}
        du\,u^{d-2}\,(u^2+y^2)^{-\frac{z+1}{2}}\Gamma(z)\,\zeta(z) \label{cint}\end{equation}
with $C >d-2 $.
We can evaluate the integral over $u$ and the remaining integral becomes

\begin{equation} 
\frac{1}{2}\int_{C-i\infty}^{C+i\infty} dz\,
y^{d-2-z}\frac{\Gamma(z)}{\Gamma(\frac{1+z}{2})}\,\Gamma(d-5/2)\,\Gamma\left(\frac{2+z-d}{2}\right)\,\zeta(z) \end{equation}
The integrand has single poles at $z=0,1$ and $z=d-2k$ where $k=1,2,3,...$.
The contour of the $z$ integration in the integral (\ref{cint}) may be closed in the
left half plane since the integrand vanishes asymptotically on the arc at infinity.
We now have to compute all the residues of the poles of the integrand.
The contribution to $N_T$ from the pole at $z=2-d$ corresponds to the
first order thermal mass $m_T^2$. From the pole at $z=1$, we get
$-T\,m/(8\pi)$ which corresponds to the sum of all the static
modes we had in the ITF (see (\ref{static})) and is nonanalytic in
the coupling ($m_T\sim \lambda_R^{1/2}$).
The sum of the contributions to $N_T$ from the poles at $z=0$ and $z=d-4$ is
\begin{equation}-\frac{m^2}{16\pi^2}\left(\log
\frac{m}{4\pi\,T}+\gamma-\frac{1}{2}\right) \end{equation} and this is
nonanalytic in the coupling as well due to the term in $\log m$.
Finally, all the remaining poles have residues of the same form and the
complete expression for $N_T$ reads (when $d=4$)
\begin{eqnarray}
\nonumber\frac{4\pi^2}{T^2}N_T(m^2)&=& \frac{\pi^2}{6}-\frac{\pi \,m}{2T}
-\frac{1}{4}\left (\frac{m}{T}\right )^2\left[\log\frac{m}{4\pi T}+\gamma-\frac{1}{2}\right] \\
&-&\frac{1}{4}\left (\frac{m}{T}\right )^2\sum_{n=1}^\infty (-1)^n 
\frac{(2n)!}{(n+1)!\,n!}\,\zeta(2n+1)
\left(\frac{m}{4\pi T}\right)^{2n}\label{exNT}\end{eqnarray}
From this expression, we see that $N_T(\delta m_{R}^2)$ contains a term involving $\log \lambda_R$ and
another one proportional to $\lambda_R^{1/2}$ . The fact that an
expansion of $N_T(\delta m_{R}^2)$ in power series of the coupling is
illegal is quite clear from (\ref{NT}) : if we try to
expand the integrand in power series of $m^2$, the
expansion parameter is actually  $m^2/p^2$ and this obviously does not
make sense since the momentum integration runs down to $0$.
If we do it anyway, we obtain what we got in perturbation
theory and we get an isolated $1/(d-4)$ term at order $\lambda_R^2$.  
Moreover, we also have 
\begin{equation} \tilde{M}(\mu^2,\delta m_R^2)=\frac{\delta
m_R^2}{32\pi^2}\,\left(\log \frac{\delta m_R^2}{\mu^2}-1\right) + {\cal O}(d-4) \end{equation}
So we find 
\begin{eqnarray}\nonumber \delta m_{R}^2 &=&
\lambda_R\,T^2\left\{\frac{1}{24}-\frac{\delta m_{R}}{8\pi\,T}
-\frac{1}{16\pi^2}\left (\frac{\delta m_{R}}{T}\right )^2\left[\log\frac{\mu}{4\pi T}+\gamma\right] \right.\\
 &-&\left.\frac{1}{16\pi^2}\left(\frac{\delta m_{R}}{T}\right )^2\sum_{n=1}^\infty (-1)^n 
\frac{(2n)!}{(n+1)!\,n!}\,\zeta(2n+1)
\left(\frac{\delta m_{R}}{4\pi T}\right)^{2n}\right\}\label{thmeq}\end{eqnarray}
One can solve this equation iteratively to
any desired accuracy in $\lambda_R$ and the first few terms in the
expansion are \cite{Landshoff}  
\begin{eqnarray} \nonumber
\frac{\delta m_R^2}{T^2}=\frac{\lambda_R}{24}&-&\frac{\lambda_R^{3/2}}{16\pi \sqrt{6}}
+\left(3-\gamma-\log\frac{\mu}{4\pi T}\right)\frac{\lambda_R^2}{384\pi^2}
\\&-&\left(1-2\gamma-2\log\frac{\mu}{4\pi T}\right)\frac{\lambda_R^{5/2}}{1024\pi^3\sqrt{2/3}}
+{\cal O}(\lambda_R^3) \label{thmexp}\end{eqnarray}
It is evident that $\delta m_R^2$ does not have an expansion in powers
of $\lambda_R$, but rather in powers of $\lambda_R^{1/2}$.
\subsection{Renormalised perturbative series}
We express $\delta m_R^2$ in terms of the renormalised coupling and
the renormalisation scale. From (\ref{ren}) and (\ref{scheme}) 
\begin{equation}
\lambda=\lambda_R\,\mu^{4-d}\left[\sum_{k=0}^{\infty}\,(-)^k
\mu^{k(4-d)}\,\lambda_R^k\,\left(M'(\mu^2)\right)^k\right] \end {equation} 
so diagrams corresponding to different fixed orders in $\lambda$ will yield
contributions both to the same order in $\lambda_R$ and to higher orders.
Evidently, to lowest order, we have $\lambda=\lambda_R$ so $\delta
m_R^2=\lambda_R T^2 /24$. To order $\lambda_R^2$, both the one and
two-loop diagrams contribute and the $1/(d-4)$ divergence coming from
$\Pi^{(1)}$ exactly cancels the one that results from the
renormalisation of the coupling in the one-loop diagram
$\Pi^{(0)}$. We are therefore left with a finite expression up to order $\lambda_R^2$:
\begin{equation} \delta m_R^2=\frac{\lambda_R\,T^2}{24}
-\frac{\lambda_R^2\,T^2}{384\pi^2}\left[\log\frac{\mu}{4\pi
T}+\gamma\right] \end{equation}
Having obtained an exact expression for $\delta m_R^2$ in terms of
$\lambda_R$, we investigate the problems associated with first
expanding it as a perturbation series in $\lambda_R$.
Comparing with (\ref{thmexp}), we see that there are missing terms.
This is because we also need to
include the contribution from the non-vanishing sum of all the static
modes : this contribution, which also yields all the terms proportional to
an odd power of $\lambda_R^{1/2}$ in (\ref{thmexp}), is missed out in the perturbative expansion
and only appears when we resumm the series.\\ 
After tedious calculations, divergences cancel out again at order $\lambda_R^3$ if we take into account all the diagrams that
contribute to the thermal mass at this order (they are represented in
figure 2): diagram D in figure 2, which is not a ring diagram, rescues the
situation by cancelling divergences coming from diagrams A and B and
the renormalisation of the bare coupling in these diagrams. It means
that, if we only consider ring diagrams, we get uncancelled
divergences from order $\lambda_R^3$ onwards in the renormalised
perturbation series. 
\begin{figure}[h] 
\epsfig{file=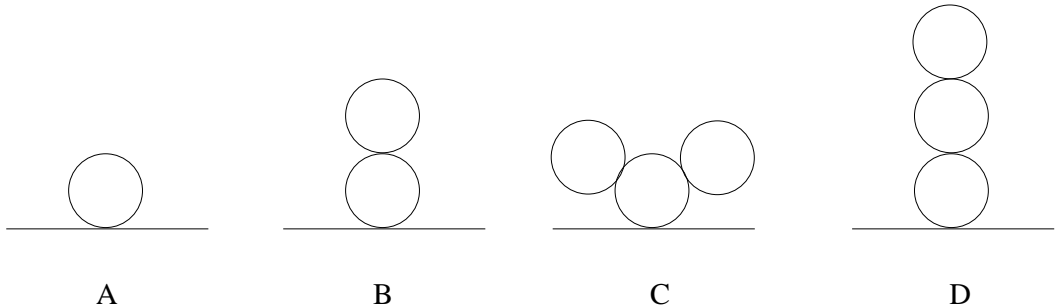,width=14cm}
\caption{Diagrams contributing to $\delta m_{R}^2$ at order $\lambda_R^3$}
\end{figure}
From order $\lambda_R^{4}$ onwards, it is not possible to
get rid of divergences any more even if we sum up all diagrams that
contribute to the thermal mass at this order : worse and worse
divergences crop up in the renormalised perturbative series when going
to higher orders.
For instance, at order $\lambda_R^{4}$, we have 
\begin{equation} \left. \delta
m_R^2\right|_{{\cal O}(\lambda_R^4)}=\frac{-\lambda_R^4\,T^2}{49152\,\pi^6(d-4)^3 } + {\cal O}((d-4)^{-2})  \end{equation}
In conclusion, we have seen in this section that the nonanalyticity of
the thermal mass in the coupling causes the failure of the corresponding Feynman perturbative
expansion. This failure has two consequences : firstly, infrared
divergences appear in the renormalised series from order
$\lambda_R^{4}$ and, secondly, the term proportional to
$\lambda^{1/2}$ and its contributions to all order in $\lambda_R$ are
missed out.     

\section{Conclusion}
\setcounter{equation}{0}

The equation (\ref{renf}) derived for the pressure is exact, or non
perturbative. Its right-hand side involves $\delta m_R^4$ and we have
seen in the previous section that we get uncancelled $1/(d-4)$
infrared singularities if we expand it in powers of $\lambda_R$ before
we take the $d\rightarrow4$ limit.
On the other hand, if we take the limit $d\rightarrow 4$ on the exact
expression (\ref{NT}), then we get an expression which cannot be
expanded in power series of the coupling any more since a 
coefficient in the expansion would be infinite. It tells us that we
have to use the more involved expansion outlined in section 4.                   
In fact, before we take the limit $d\rightarrow 4$, nothing prevents
us from expanding in power series of the coupling even if this
expansion is illegal in four space-time dimensions.
We could therefore conclude by saying that the correct procedure is
to work in four space-time dimensions before expanding the physical 
quantity under consideration in a suitable way.
In this paper, we studied both the perturbative and the exact expression 
of the finite temperature contribution to the thermal mass and we saw
that the infrared divergent $\lambda^2/(d-4)$ term in the unresummed
perturbative series is effectively replaced with $\lambda\,\log
\lambda$ in the corresponding exact expansion (see section 4.2).
The same kind of mechanism may be related to the severe infrared
problems encountered in thermal QCD. In the case of the QCD thermal
pressure, it is known that diagrams with
magnetostatic gluon loops become divergent from order $g^6$ in
perturbation theory. From what we saw in this work in the case of
large-$N$ scalar theory, it seems sensible to assume that this
$g^6/(d-4)$ divergence in the perturbative QCD pressure corresponds to
a term involving $g^6\log g$ in the exact expression for the pressure.
This will be the subject of subsequent work.    

\section*{Acknowledgments}
I am indebted to P.V.Landshoff for reading the manuscript and for
invaluable discussions. The work is supported by the Cambridge European
Trust, Christ's College and by the Department of Applied Mathematics
and Theoretical Physics.

\appendix
\section{Appendix}

\renewcommand{\theequation}{A.\arabic{equation}}
\subsection{Ring diagrams in the real-time formalism}

In the real-time formalism, the self-energy also has a matrix
structure and the matrix element relevant to the evaluation of
$\Pi^{(1)}(d,T)$ is 
\begin{equation} \Pi_{11}^{(1)}(d,T)=-i\frac{\lambda}{2}\;M_T^2\int\frac{d^{d}p}{(2\pi)^{d}}\left[D_{11}^2(p)-D_{12}^2(p) D_{21}^2(p)\right]             \end{equation}
Using (\ref{mstruct}), we get
\begin{equation} D_{11}^2(p)-D_{12}^2(p)
D_{21}^2(p)=\left(\frac{i}{p^2+i\epsilon}\right)^2-n(p^0)\left[\frac{1}{(p^2+i\epsilon)^2}-\frac{1}{(p^2-i\epsilon)^2}\right] \end{equation}
The finite temperature part of the latter expression can be written as
\begin{equation} \left.-n(p^0) \frac{\partial}{\partial
m^2}\left(\frac{1}{p^2-m^2+i\epsilon}-\frac{1}{p^2-m^2-i\epsilon}\right)\right|_{m^2=0}
\label{derivm2}\end{equation}
which is the derivative of the one-loop diagram for the thermal mass.
We therefore obtain
\begin{equation}
\int\frac{d^{d}p}{(2\pi)^{d}}\left[D_{11}^2(p)-D_{12}^2(p)
D_{21}^2(p)\right]=-i\pi(d-3)\int\frac{d\Omega_{d-1}}{(2\pi)^d}\int_0^{\infty}dp\,\frac{p^{d-5}}{e^{\beta
p}-1} \label{IRdiv}\end{equation}
Finally, if we use the representation
\begin{equation} \zeta(z)=\frac{1}{\Gamma(z)}\int_0^{+\infty}du
\frac{u^{z-1}}{e^u-1} \end{equation} and the relation
\begin{equation}\zeta(1-z)=2\,(2\pi)^{-z}\cos\left(\frac{z\pi}{2}\right)\,\Gamma(z)\zeta(z) \end{equation}
we obtain the same result (\ref{pi1}) as with the imaginary-time formalism.
Since the momentum integral in the right-hand side of (\ref{IRdiv}) is
infrared divergent in four space-time dimensions, it tells us that the corresponding $1/(d-4)$
divergence really is an infrared one.
Similarly, we could also rederive the general expression for a ring
diagram with an arbitrary number of external loops and check that it
agrees with what we found using the imaginary-time formalism : one
just has to take higher derivatives with respect to $m^2$ in
(\ref{derivm2}).\\

\subsection{The Mellin summation formula}

The Mellin transform of a function $f(x)$ is usually defined by
\begin{equation}
\phi(z)=\int_0^{\infty} x^{z-1}\,f(x)\,dx \end{equation}
The transform $\phi(z)$ exists if the integral
\begin{equation}\int_0^{\infty} \left|f(x)\right|\,x^{k-1}\,dx \end{equation}
is bounded for some $k>0$, in which case the inverse transform is
\begin{equation}
f(x)=\frac{1}{2i\pi}\int_{C-i\infty}^{C+i\infty}\,x^{-z}\phi(z)\,dz \end{equation}
where $C>k$.
Together with the standard definition of the
Riemann Zeta-function, this gives the summation formula 
\begin{equation}\sum_{n=1}^{\infty}
f(n)=\frac{1}{2i\pi}\,\int_{C-i\infty}^{C+i\infty} dz\,
\zeta(z)\,\phi(z) \end{equation}

\end{document}